\documentclass[twocolumn,aps,showpacs,superscriptaddress]{revtex4-1}
\usepackage{amsmath}
\usepackage{graphicx}
\usepackage[none]{hyphenat}
\RequirePackage[pagewise,mathlines]{lineno}

\begin{document}
\title{Observation of excess J/$\psi$ yield at very low transverse momenta in Au+Au collisions at $\sqrt{s_{\rm{NN}}} =$ 200 GeV and U+U collisions at $\sqrt{s_{\rm{NN}}} =$ 193 GeV}%
\affiliation{Abilene Christian University, Abilene, Texas   79699}
\affiliation{AGH University of Science and Technology, FPACS, Cracow 30-059, Poland}
\affiliation{Alikhanov Institute for Theoretical and Experimental Physics, Moscow 117218, Russia}
\affiliation{Argonne National Laboratory, Argonne, Illinois 60439}
\affiliation{Brookhaven National Laboratory, Upton, New York 11973}
\affiliation{University of California, Berkeley, California 94720}
\affiliation{University of California, Davis, California 95616}
\affiliation{University of California, Los Angeles, California 90095}
\affiliation{University of California, Riverside, California 92521}
\affiliation{Central China Normal University, Wuhan, Hubei 430079 }
\affiliation{University of Illinois at Chicago, Chicago, Illinois 60607}
\affiliation{Creighton University, Omaha, Nebraska 68178}
\affiliation{Czech Technical University in Prague, FNSPE, Prague 115 19, Czech Republic}
\affiliation{Technische Universit\"at Darmstadt, Darmstadt 64289, Germany}
\affiliation{E\"otv\"os Lor\'and University, Budapest, Hungary H-1117}
\affiliation{Frankfurt Institute for Advanced Studies FIAS, Frankfurt 60438, Germany}
\affiliation{Fudan University, Shanghai, 200433 }
\affiliation{University of Heidelberg, Heidelberg 69120, Germany }
\affiliation{University of Houston, Houston, Texas 77204}
\affiliation{Huzhou University, China}
\affiliation{Indiana University, Bloomington, Indiana 47408}
\affiliation{Institute of Physics, Bhubaneswar 751005, India}
\affiliation{University of Jammu, Jammu 180001, India}
\affiliation{Joint Institute for Nuclear Research, Dubna 141 980, Russia}
\affiliation{Kent State University, Kent, Ohio 44242}
\affiliation{University of Kentucky, Lexington, Kentucky 40506-0055}
\affiliation{Lawrence Berkeley National Laboratory, Berkeley, California 94720}
\affiliation{Lehigh University, Bethlehem, Pennsylvania 18015}
\affiliation{Max-Planck-Institut f\"ur Physik, Munich 80805, Germany}
\affiliation{Michigan State University, East Lansing, Michigan 48824}
\affiliation{National Research Nuclear University MEPhI, Moscow 115409, Russia}
\affiliation{National Institute of Science Education and Research, HBNI, Jatni 752050, India}
\affiliation{National Cheng Kung University, Tainan 70101 }
\affiliation{Nuclear Physics Institute of the CAS, Rez 250 68, Czech Republic}
\affiliation{Ohio State University, Columbus, Ohio 43210}
\affiliation{Institute of Nuclear Physics PAN, Cracow 31-342, Poland}
\affiliation{Panjab University, Chandigarh 160014, India}
\affiliation{Pennsylvania State University, University Park, Pennsylvania 16802}
\affiliation{NRC "Kurchatov Institute", Institute of High Energy Physics, Protvino 142281, Russia}
\affiliation{Purdue University, West Lafayette, Indiana 47907}
\affiliation{Pusan National University, Pusan 46241, Korea}
\affiliation{Rice University, Houston, Texas 77251}
\affiliation{Rutgers University, Piscataway, New Jersey 08854}
\affiliation{Universidade de S\~ao Paulo, S\~ao Paulo, Brazil 05314-970}
\affiliation{University of Science and Technology of China, Hefei, Anhui 230026}
\affiliation{Shandong University, Qingdao, Shandong 266237}
\affiliation{Shanghai Institute of Applied Physics, Chinese Academy of Sciences, Shanghai 201800}
\affiliation{Southern Connecticut State University, New Haven, Connecticut 06515}
\affiliation{State University of New York, Stony Brook, New York 11794}
\affiliation{Temple University, Philadelphia, Pennsylvania 19122}
\affiliation{Texas A\&M University, College Station, Texas 77843}
\affiliation{University of Texas, Austin, Texas 78712}
\affiliation{Tsinghua University, Beijing 100084}
\affiliation{University of Tsukuba, Tsukuba, Ibaraki 305-8571, Japan}
\affiliation{United States Naval Academy, Annapolis, Maryland 21402}
\affiliation{Valparaiso University, Valparaiso, Indiana 46383}
\affiliation{Variable Energy Cyclotron Centre, Kolkata 700064, India}
\affiliation{Warsaw University of Technology, Warsaw 00-661, Poland}
\affiliation{Wayne State University, Detroit, Michigan 48201}
\affiliation{Yale University, New Haven, Connecticut 06520}

\author{J.~Adam}\affiliation{Creighton University, Omaha, Nebraska 68178}
\author{L.~Adamczyk}\affiliation{AGH University of Science and Technology, FPACS, Cracow 30-059, Poland}
\author{J.~R.~Adams}\affiliation{Ohio State University, Columbus, Ohio 43210}
\author{J.~K.~Adkins}\affiliation{University of Kentucky, Lexington, Kentucky 40506-0055}
\author{G.~Agakishiev}\affiliation{Joint Institute for Nuclear Research, Dubna 141 980, Russia}
\author{M.~M.~Aggarwal}\affiliation{Panjab University, Chandigarh 160014, India}
\author{Z.~Ahammed}\affiliation{Variable Energy Cyclotron Centre, Kolkata 700064, India}
\author{I.~Alekseev}\affiliation{Alikhanov Institute for Theoretical and Experimental Physics, Moscow 117218, Russia}\affiliation{National Research Nuclear University MEPhI, Moscow 115409, Russia}
\author{D.~M.~Anderson}\affiliation{Texas A\&M University, College Station, Texas 77843}
\author{R.~Aoyama}\affiliation{University of Tsukuba, Tsukuba, Ibaraki 305-8571, Japan}
% ...

\collaboration{STAR Collaboration}\noaffiliation
\date{\today}%
\begin{abstract}
We report on the first measurements of J/$\psi$ production at very low transverse momentum ($p_{T} <$ 0.2 GeV/c) in hadronic Au+Au collisions at $\sqrt{s_{\rm{NN}}} =$ 200 GeV and U+U collisions at $\sqrt{s_{\rm{NN}}} =$ 193 GeV. Remarkably, the inferred nuclear modification factor of J/$\psi$ at mid-rapidity in Au+Au (U+U) collisions reaches about 24 (52) for $p_{T} <$ 0.05 GeV/c in the 60-80$\%$ collision centrality class. This noteworthy enhancement cannot be explained by hadronic production accompanied by cold and hot medium effects. In addition, the $dN/dt$ distribution of J/$\psi$ for the very low $p_{T}$ range is presented for the first time. The distribution is consistent with that expected from the Au nucleus and shows a hint of interference. Comparison of the measurements to theoretical calculations of coherent production shows that the excess yield can be described reasonably well and reveals a partial disruption of coherent production in semi-central collisions, perhaps due to the violent hadronic interactions. Incorporating theoretical calculations, the results strongly suggest that the dramatic enhancement of J/$\psi$ yield observed at extremely low $p_{T}$ originates from coherent photon-nucleus interactions. In particular, coherently produced J/$\psi$'s in violent hadronic collisions may provide a novel probe of the quark-gluon-plasma.
\end{abstract}
\maketitle
In ultra-relativistic heavy-ion collisions, one aims to examine the properties of a new form of matter - the quark-gluon plasma (QGP), which was predicted using lattice Quantum Chromodynamics (QCD) calculations~\cite{PBM_QGP}, and study its properties in the laboratory. J/$\psi$ suppression in heavy-ion collisions  has been  proposed as a ``smoking gun'' of QGP formation~\cite{MATSUI1986416} owing to the color screening effect in the deconfined medium. Over the past twenty years, various measurements of J/$\psi$ production in heavy-ion collisions have been carried out in different collision systems and at different energies~\cite{PhysRevLett.99.132302,Abreu1999408,Abreu200028,PhysRevC.90.024906,PhysRevLett.98.232301,PhysRevLett.109.072301,Aad2011294,Chatrchyan2012}. The interpretation of these observations impelled an introduction of a regeneration effect (recombination of charm quarks in the QGP)~\cite{PhysRevLett.97.232301} and Cold Nuclear Matter (CNM; nuclear shadowing, initial energy loss, cronin, etc.) effects~\cite{FERREIRO200950} to the J/$\psi$ modification in heavy-ion collisions. At present, the interplay of color screening, regeneration, and CNM effects can reasonably well describe the J/$\psi$ suppression at SPS, RHIC, and LHC energies in heavy-ion collisions~\cite{Adamczyk201713}.

The strong electromagnetic fields generated by the colliding ions can be represented by a spectrum of equivalent photons~\cite{EPA1,EPA2};  therefore heavy-ion collisions can be used to study coherent photonuclear interactions~\cite{UPCreview}. J/$\psi$ can be produced in photon-nucleus interactions via Pomeron-exchange, the perturbative-QCD equivalent of which is the exchange of two gluons or a gluon ladder~\cite{PhysRevD.57.512}. Coherently produced J/$\psi$s in heavy-ion collisions are expected to probe the nuclear gluon-distribution at low Bjorken-$x$~\cite{REBYAKOVA2012647}, for which there is still considerable uncertainty~\cite{1126-6708-2009-04-065}. The coherent nature of the interactions leads to a distinctive configuration; the final products consist of two intact nuclei and only a J/$\psi$ with very low transverse momentum ($p_{T} <$ 0.1 GeV/c). Conventionally, the products of these reactions are only detectable when there are no accompanying hadronic interactions, i.e. in the so-called Ultra-Peripheral Collisions (UPC) where the impact parameter ($b$) is larger than twice the nuclear radius ($R_{A}$). Several results of J/$\psi$ production in UPC are already available at RHIC~\cite{Afanasiev2009321} and LHC~\cite{20131273,Abbas2013,2017489}, which provide valuable insights into the gluon distribution in the colliding nuclei~\cite{Guzey:2013qza}.

Can the coherent photonuclear interaction also occur in hadronic heavy-ion collisions ($b < 2R_{A}$), where the nuclei collide and break up? Recently, a significant excess of J/$\psi$ yield at very low $p_{T}$ ($<$ 0.3 GeV/c) has been observed by the ALICE Collaboration in peripheral hadronic Pb+Pb collisions at $\sqrt{s_{\rm{NN}}} =$ 2.76 TeV at forward-rapidity~\cite{LOW_ALICE}. It cannot be explained by the scenario of hadronic production modified by color screening, regeneration, and CNM effects. The observed excess may originate from coherent photoproduction, which imposes great challenges for the existing coherent photoproduction models, e.g., how the broken nuclei satisfy the requirement of coherence. Measurements of J/$\psi$ production at very low $p_{T}$ at different collision energies, collision systems, and centralities can shed new light on the origin of the excess.

In this letter, the first RHIC results on J/$\psi$ production at very low $p_{T}$ in hadronic heavy-ion collisions are presented. J/$\psi$ production yields in Au+Au collisions at $\sqrt{s_{\rm{NN}}} =$ 200 GeV and U+U collisions at $\sqrt{s_{\rm{NN}}} =$ 193 GeV are measured at mid-rapidity via the dielectron decay channel. Significant enhancement of J/$\psi$ production at very low $p_{T}$ has been observed with respect to expectations from hadroproduction. Furthermore, the excess yield is studied as functions of centrality, transverse momentum, and collision system and is compared to model calculations incorporating the coherent photoproduction scenario.

The STAR experiment is a large-acceptance multi-purpose detector which covers the full azimuth in the pseudorapidity interval of $|\eta|<1$~\cite{ACKERMANN2003624}. The Vertex Position Detector (VPD)~\cite{LLOPE201423}, which is located at $4.24<|\eta|<5.1$, was used to select collisions that were within $\pm$ 30 cm of the center of the STAR detector along the beam direction. The minimum-bias trigger used in this analysis requires a coincidence between the East and West VPD. The Au+Au data at $\sqrt{s_{\rm{NN}}}=$ 200 GeV were collected during the 2010 and 2011 RHIC runs, while the U+U data at $\sqrt{s_{\rm{NN}}} =$ 193 GeV were collected in 2012. The total numbers of events used in Au+Au and U+U collisions are  720 million and 270 million, respectively. The collision centrality is determined by comparing the measured charged particle multiplicity within $|\eta| < 0.5$ with a Monte Carlo Glauber model simulation~\cite{doi:10.1146/annurev.nucl.57.090506.123020}. The effects of acceptance and efficiency changes on the measured $dN/d\eta$ due to the luminosity and collision vertex variations have been taken into account. In order to avoid the significant inefficiency of the VPD in peripheral collisions, only data in 0-80$\%$ central collisions are accepted. For each collision centrality, an average nuclear overlap function, $\langle T_{\rm{AA}} \rangle$, average number of participants, $\langle N_{\rm{part}} \rangle$, and average number of binary collisions, $\langle N_{\rm{coll}} \rangle$, are calculated based on the Glauber model. The primary sub-detectors used in this analysis include the Time Projection Chamber (TPC) ~\cite{ANDERSON2003659}, the Time-of-Flight (TOF) detector ~\cite{LLOPE2012S110}, and the Barrel Electromagnetic Calorimeter (BEMC) ~\cite{BEDDO2003725}. The TPC provides tracking and particle identification via the ionization energy loss ($\langle dE/dx\rangle$) of charged particles. The TOF ~\cite{LLOPE2012S110} measures the velocity of particles, which greatly improved electron identification at low momenta. The BEMC ~\cite{BEDDO2003725}, a lead-scintillator calorimeter, is used to improve electron identification at relative high momenta ($p > 1.5$ GeV/$c$).

\renewcommand{\floatpagefraction}{0.75}
\begin{figure}[htbp]
\begin{center}
\includegraphics[keepaspectratio,width=0.5\textwidth]{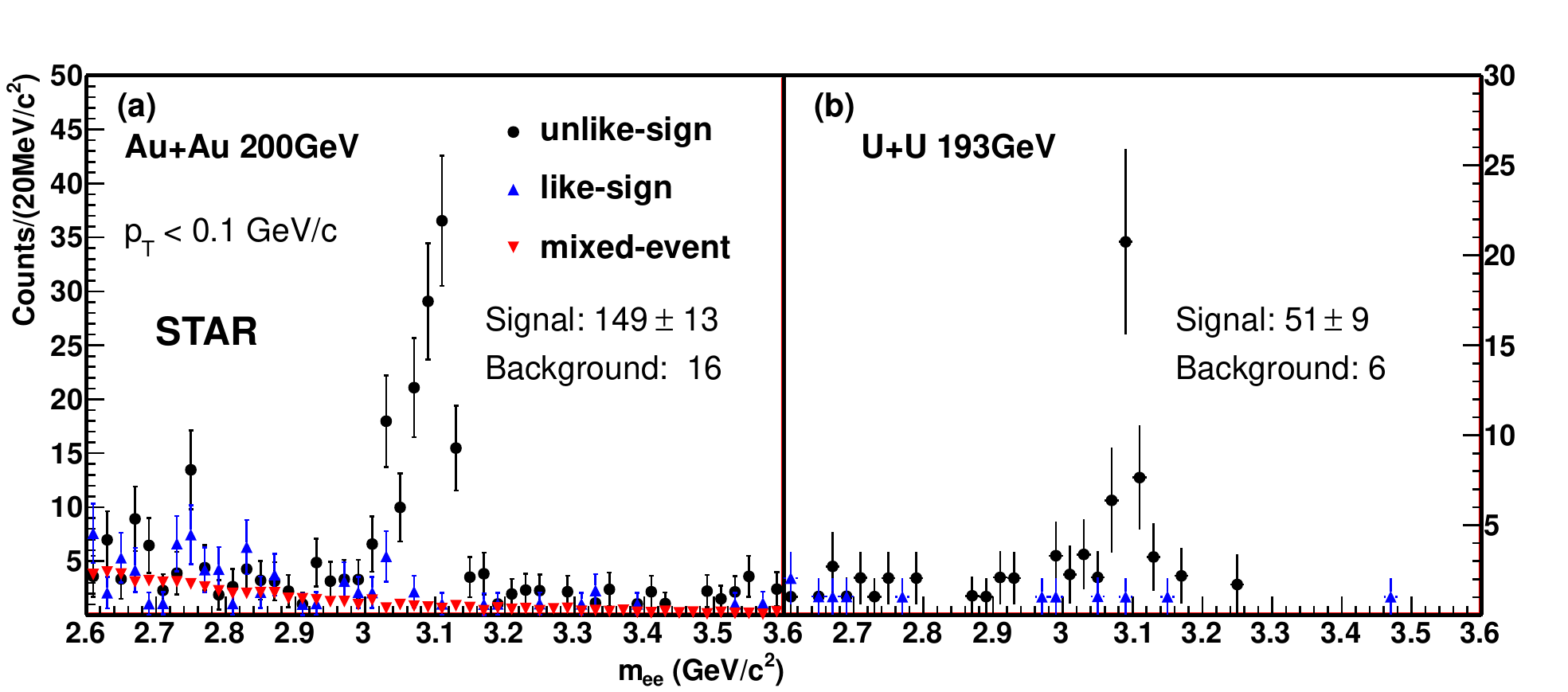}
\caption{(color online) The dielectron invariant mass spectrum for the 40-80$\%$ centrality class in Au+Au collisions at $\sqrt{s_{\rm{NN}}} =$ 200 GeV (a) and in U+U collisions at $\sqrt{s_{\rm{NN}}} =$ 193 GeV (b). The error bars are the statistical uncertainties.}
\label{figure1}
\end{center}
\end{figure}

In this analysis, the J/$\psi$s are reconstructed through their decay into electron-positron pairs, $\text{J}/\psi \rightarrow e^{+}e^{-}$ (branching ratio Br($J/\psi \rightarrow e^{+}e^{-}$) = 5.97$\pm$0.03$\%$~\cite{1674-1137-40-10-100001}). The daughter tracks are required to have at least 25 TPC hits, and a distance of closest approach (DCA) to the primary vertex less than 3 cm for $p < 1.5$ GeV/$c$ and $< 1$ cm for $p > 1.5$ GeV/$c$. The electron and positron candidates are identified by their specific energy loss ($\langle dE/dx\rangle$) in the TPC. More than 15 TPC hits were required to calculate $\langle dE/dx\rangle$. Electron and positron candidates are further separated from hadrons by selecting on the inverse velocity $1/\beta$, where $\beta$ is the velocity determined from TOF information and normalized by the speed of light. In Au+Au collisions, a cut on the ratio of momentum to energy deposited in the BEMC is used to further suppress hadrons for high momentum candidates. The combination of these cuts enables the identification of electrons and positrons over a wide momentum range~\cite{PhysRevC.90.024906,201355,Adamczyk201713}. The electron sample purity integrated over the measured momentum region is over 90$\%$. The J/$\psi$ measurements cover the rapidity range $|y|<1$ due to the STAR acceptance and decay kinematics.

The J/$\psi$ candidates are reconstructed by combining pairs of electron-positron candidates with $p_{T} \ge$ 0.2 GeV/c and $|\eta| \le$ 1 in the same event. The combinatorial background in Au+Au collisions is estimated via the mixed-event technique~\cite{Adamczyk201713}, which could significantly reduce the statistical uncertainty in comparison to the like-sign technique. However, in U+U collisions, the like-sign technique is employed, since the mixed-event technique could not reproduce the combinatorial background well. The invariant mass distributions of $e^{+} e^{-}$ pairs in 40-80$\%$ central Au+Au collisions and U+U collisions are shown in Fig.~\ref{figure1}. The invariant mass distribution of  $e^{+}e^{-}$ pairs after combinatorial background subtraction is then fitted using the J/$\psi$ signal shape obtained from MC simulation, which includes momentum resolution, electron bremsstrahlung, and J$/\psi$ internal radiation~\cite{Spiridonov:2004mp}, combined with an exponential function for the residual background. The residual background mainly originates from the decays of correlated charm hadrons, Drell-Yan processes and possible coherent photon-photon interactions. The raw J/$\psi$ signal is obtained from bin counting in the mass range 2.9 - 3.2 GeV/$c^{2}$ after subtraction of the combinatorial background, while the residual background is assigned as a source of uncertainty. The raw counts in this mass range are $149\pm13$ for Au+Au collisions and $51\pm9$ for U+U collisions. The fraction of J/$\psi$ counts outside of the bin counting window is determined from the simulated J/$\psi$ signal shape and is found to be $\sim$ 5$\%$, which is used to correct the raw J$/\psi$ counts.

\renewcommand{\floatpagefraction}{0.75}
\begin{figure}[htbp]
\begin{center}
\includegraphics[keepaspectratio,width=0.4\textwidth]{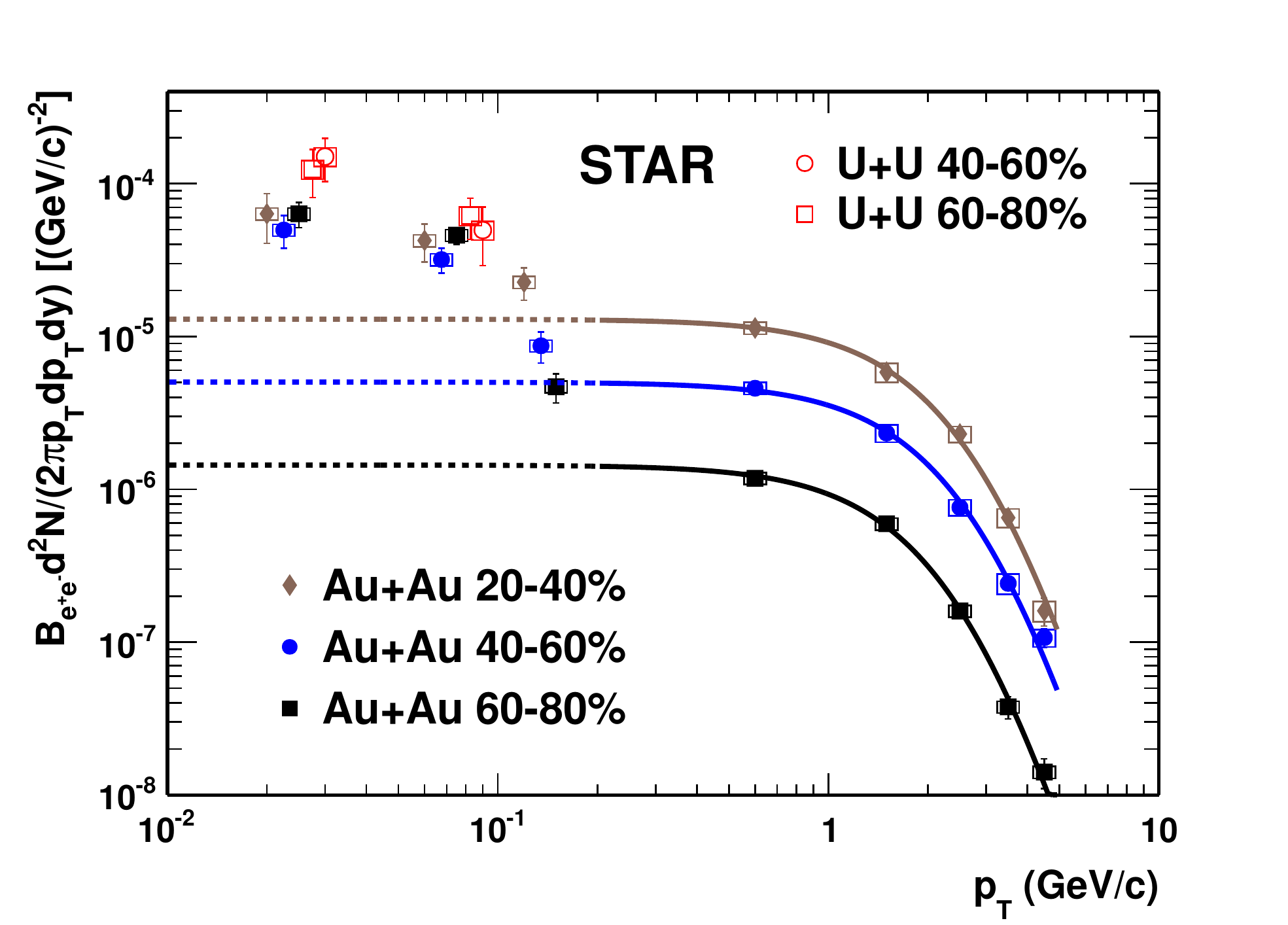}
\caption{(color online)  The J/$\psi$ invariant yields for Au+Au collisions at $\sqrt{s_{\rm{NN}}} =$  200 GeV and U+U collisions at $\sqrt{s_{\rm{NN}}} =$ 193 GeV as a function of $p_{T}$ for different centralities at mid-rapidity ($|y| <$ 1). The error bars depict the statistical errors while the boxes represent the systematic uncertainties. The data points with $p_{T} <$ 0.2 GeV/c have been slightly shifted along the horizontal axis to facilitate viewing of the data. The solid lines in the figure are the fits to data points in the range of $p_{T} >$ 0.2 GeV/c, while the dashed lines are the extrapolations of the fits.}
\label{figure2}
\end{center}
\end{figure}

The acceptance and efficiency corrections, such as TPC tracking, BEMC matching, and $p/E$ cut, are evaluated via a GEANT3~\cite{PhysRevC.79.034909} simulation of the STAR detector. Other efficiency corrections such as those corresponding to the $dE/dx$ and TOF related cuts are obtained directly from data~\cite{PhysRevC.92.024912}. The acceptance and efficiency correction procedure used is very similar to Refs.~\cite{PhysRevC.90.024906,201355,Adamczyk201713}, except that the J$/\psi$'s at very low $p_{T}$ ($p_{T} < 0.2$~GeV/c) are set to be transversely polarized to meet the coherent production requirement~\cite{UPCreview}.

In this analysis, the systematic uncertainties on the efficiency correction from the GEANT3 simulation are estimated by comparing the related cut variable distributions between simulation and data, while the systematic uncertainties on data driven efficiencies are extracted by varying electron samples with different purities. The systematic uncertainties from yield extraction are evaluated by taking the residual background contribution under the mass-counting region and changing the normalization range for mixed-events. The associated uncertainties include uncertainties from the TPC tracking (Au+Au: $\sim 4\%$; U+U: $\sim 4\%$ ), the electron identification in the TPC (Au+Au: $\sim 1\%$; U+U: $\sim 1\%$), TOF (Au+Au: $\sim 1\%$; U+U: $\sim 3\%$), and BEMC (Au+Au: $\sim 3\%$), internal radiation (Au+Au: $\sim 4\%$; U+U: $\sim 4\%$), and the yield extraction procedure (Au+Au: $\sim 6\%$; U+U: $\sim 13\%$).  The total systematic uncertainties are the quadratic sums of the individual sources (Au+Au: $\sim 9\%$; U+U: $\sim 14\%$).

Figure~\ref{figure2} shows the J/$\psi$ invariant yields for Au+Au collisions at $\sqrt{s_{\rm{NN}}} =$ 200 GeV and U+U collisions at $\sqrt{s_{\rm{NN}}} =$ 193 GeV as a function of $p_{T}$ for different centralities at mid-rapdity ($|y| <$1). It should be pointed out that the data points used in this letter with $p_{T} > $ 1~GeV/c for collision centralities 20-40$\%$ and 40-60$\%$ are from previous STAR measurements~\cite{PhysRevC.90.024906} using the same datasets. Compared with the data points at $p_{T} >$ 0.2 GeV/c, the results in the region of $p_{T} < 0.2$ GeV/c seem to follow a different trend, especially in 40-80$\%$ peripheral collisions. The solid lines in the figure are the fits to data points in the range of $p_{T} >$ 0.2~GeV/c using Eq.~\ref{equation1}:
  \begin{equation}
  \label{equation1}
  \frac{d^{2}N}{2{\pi}p_{T}dp_{T}dy} = \frac{a}{(1+b^{2}p_{T}^{2})^{n}},
  \end{equation}
where $a$, $b$, and $n$ are free parameters. This empirical functional form can well describe the world-wide $p_{T}$ spectra of J/$\psi$ both in p+p~\cite{PhysRevC.93.024919}. The extrapolations of the fits to the range of $p_{T} <$ 0.2~GeV/c, shown as dashed lines, have been made to illustrate the expected contribution of J/$\psi$ production in this $p_{T}$ range. As shown in the figure, the fits describe the data points above 0.2 GeV/c very well, but significantly underestimate the yields below 0.2 GeV/c for non-central collisions (20-80$\%$).

\renewcommand{\floatpagefraction}{0.75}
\begin{figure}[htbp]
\begin{center}
\includegraphics[keepaspectratio,width=0.4\textwidth]{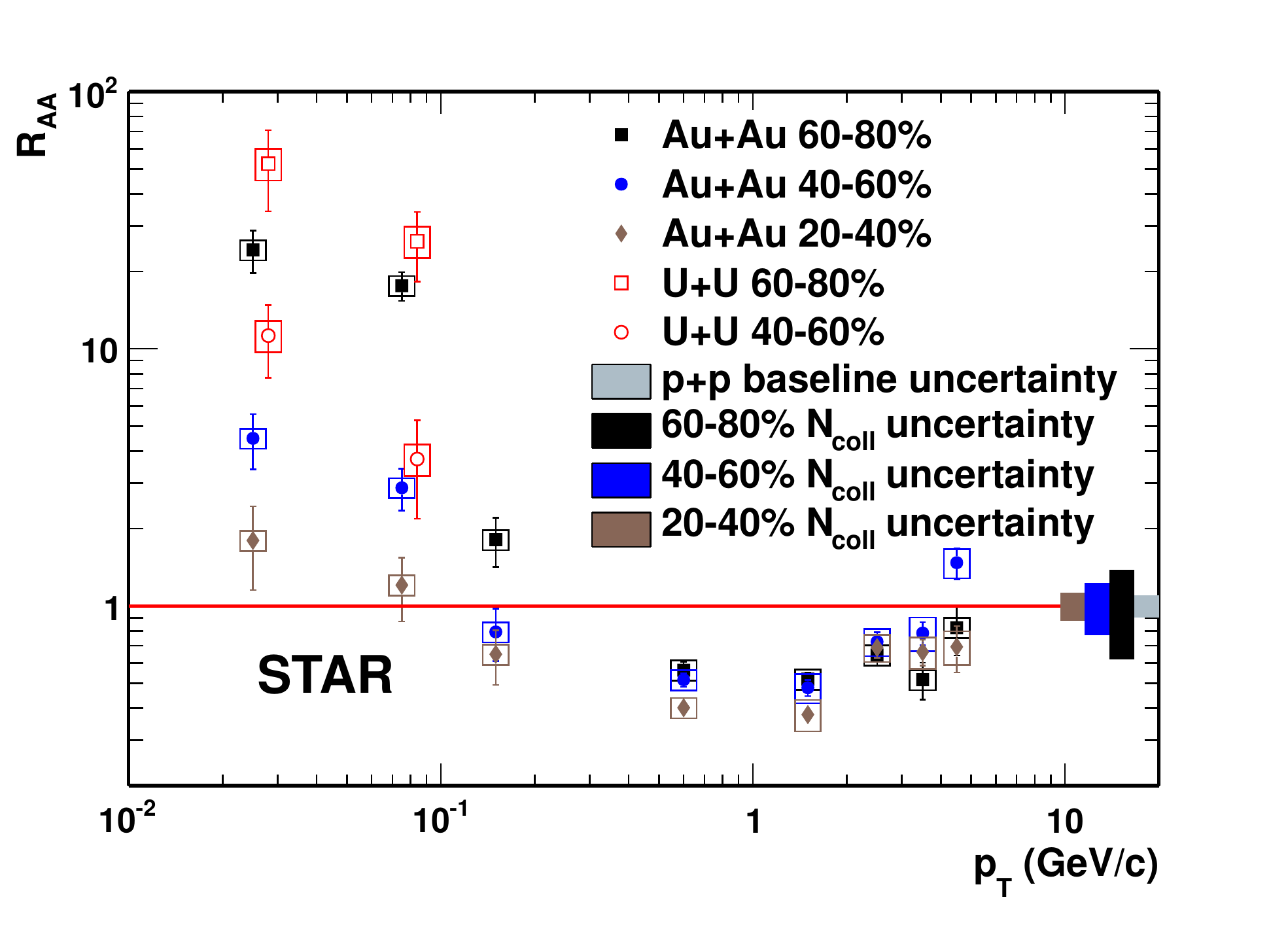}
\caption{(color online) The J/$\psi$ $R_{\rm{AA}}$ as a function $p_{T}$ in Au+Au collisions at $\sqrt{s_{\rm{NN}}} =$ 200 GeV and U+U collisions at $\sqrt{s_{\rm{NN}}} =$ 193 GeV. The error bars represent the statistical uncertainties, while the boxes represent the systematic uncertainties. The shaded bands at unity indicate the uncertainties on the p+p reference~\cite{PhysRevC.93.024919} and $\langle N_{coll} \rangle$. }
\label{figure3}
\end{center}
\end{figure}

Figure~\ref{figure3} represents the J/$\psi$ nuclear modification factor ($R_{\rm{AA}}$) as a function of $p_{T}$ in Au+Au collisions and U+U collisions for different centrality classes. The p+p baseline of $R_{\rm{AA}}$ estimation for $0 < p_T < 0.2$ GeV/c is derived by the approach described in Ref.~\cite{PhysRevC.93.024919} using the world-wide experimental data, since there is no measurement at $\sqrt{s} =$ 200 GeV. Suppression of J/$\psi$ production is observed for $p_{T} >$ 0.2 GeV/c in all collision centrality classes, which is consistent with the previous measurements~\cite{PhysRevC.90.024906,201355,Adamczyk201713,PhysRevLett.98.249902} and can be well described by the transport models~\cite{PhysRevC.82.064905,LIU200972} incorporating cold and hot medium effects. However, in the extremely low $p_{T}$ range , i.e., $p_T < 0.2$ GeV/c, a large enhancement of $R_{\rm{AA}}$ above unity is observed in peripheral collisions (40-80$\%$) both for Au+Au and U+U collisions. In this $p_{T}$ range, the color screening and CNM effects would suppress J/$\psi$ production, and the only gain effect, which is regeneration, is negligible in peripheral collisions~\cite{LIU200972}. The overall effect would lead to $R_{\rm{AA}} <$ 1 for hadronic production, which is far below the current measurement. For $p_T < 0.05$ GeV/c in the 60-80$\%$ centrality class, the $R_{\rm{AA}}$ is $24\pm5\pm9 (\rm{syst.})$ for Au+Au collisions and $52\pm18\pm16 (\rm{syst.})$ for U+U collisions, strongly suggesting that an additional production mechanism other than hadronic production is responsible for the observed yield at very low $p_{T}$.

\renewcommand{\floatpagefraction}{0.75}
\begin{figure}[htbp]
\begin{center}
\includegraphics[keepaspectratio,width=0.4\textwidth]{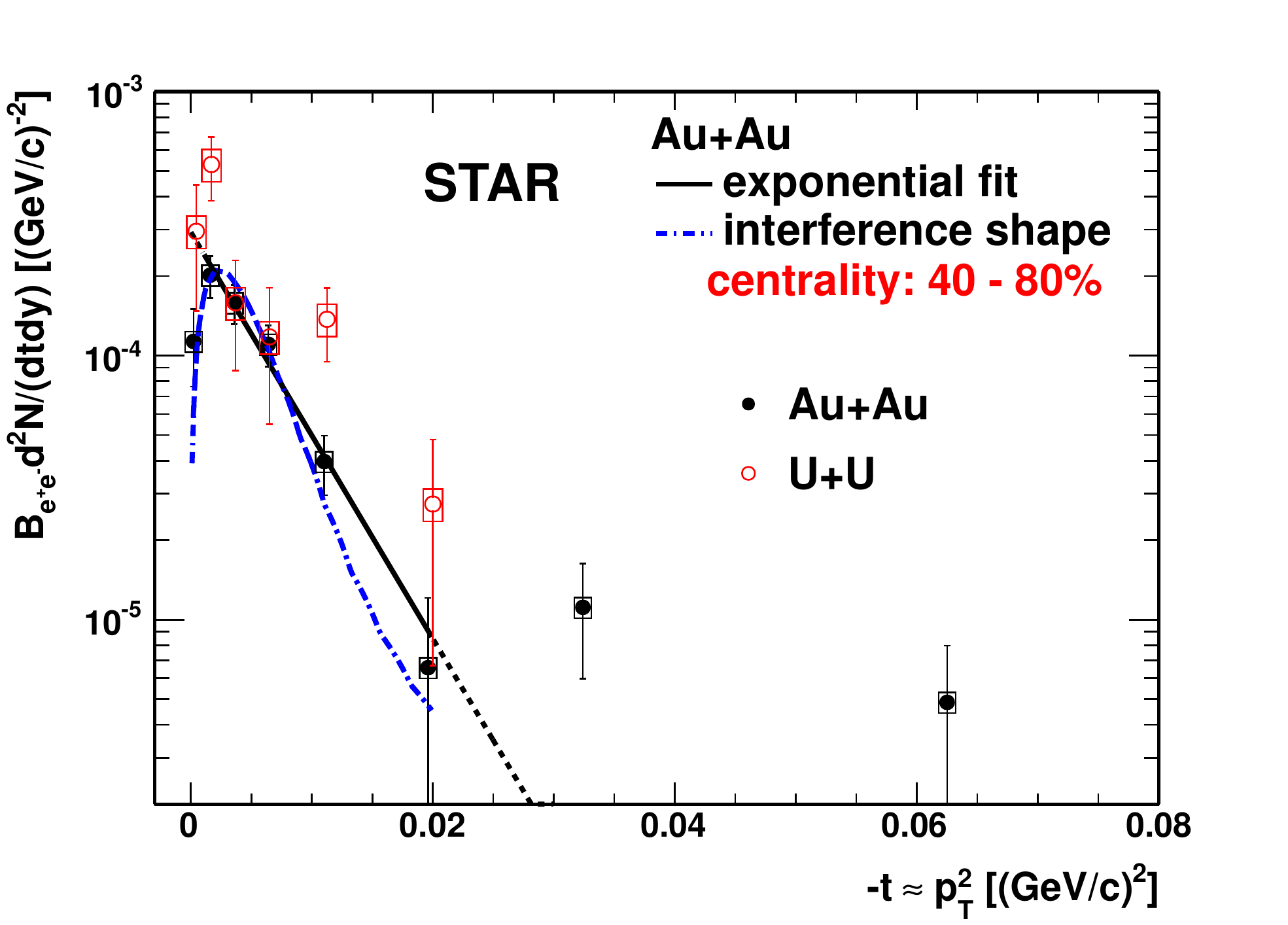}
\caption{(color online) The J/$\psi$ yield with the expected hadronic contribution subtracted as a function of the negative momentum transfer squared $-t$ ( $-t \sim p_{T}^{2}$ ) for the 40-80$\%$ collision centrality class in Au+Au and U+U collisions. The error bars represent the statistical uncertainties, while the boxes represent the systematic uncertainties. The black solid line is an exponential fit to the Au+Au data points in the range of 0.002-0.015 $(\text{GeV/c})^{2}$. The dashed black lines are extrapolations of the fit. The blue dash-dotted line is a fit to the Au+Au data points in the range of 0-0.015 $(\text{GeV/c})^{2}$ with the interference shape from~\cite{PhysRevC.97.044910}.}
\label{figure4}
\end{center}
\end{figure}

Considering the fact that the excess is observed in the extremely low $p_T$ region, a plausible scenario for the additional production mechanism is coherent photoproduction. Coherent photoproduction has been studied in detail for UPC in heavy-ion collisions~\cite{Afanasiev2009321,20131273,Abbas2013}. The differential cross section $d\sigma/dt$ for coherent products is a key measurement in UPC. It reveals the distribution of interaction sites and is closely related to the parton distribution in the nucleus. The Mandelstam variable $t \simeq -p_{T}^{2}$ at RHIC top energy. Figure~\ref{figure4} shows the J/$\psi$ yield with the expected hadronic contribution subtracted as a function of $-t$ for the 40-80$\%$ centrality class in Au+Au and U+U collisions in the low $p_{T}$ range. The expected hadronic contributions are extracted from the fit extrapolations shown in Fig.~\ref{figure2}. In order to assess systematic uncertainties, the following parametrization of J/$\psi$ production from hadronic contribution  as a function of $p_{T}$ in a given centrality class has been used:
  \begin{equation}
  \label{eqa_1}
  \frac{dN^{h}_{\rm{AA}}}{dp_{T}} = \langle T_{\rm{AA}} \rangle \times \frac{d\sigma^{\text{J}/\psi}_{pp}}{dp_{T}} \times R_{\rm{AA}}^{\text{J}/\psi_{h}},
  \end{equation}
  where $R_{\rm{AA}}^{\text{J}/\psi_{h}}$ is given by the transport model calculations~\cite{LIU200972,PhysRevC.82.064905}.
The shape of the $dN/dt$ distribution is very similar to that observed in UPC~\cite{PhysRevC.77.034910}. An exponential fit has been applied to the distribution in the  $-t$ range of 0.001-0.015 $(\text{GeV/c})^{2}$ for Au+Au collisions. The slope parameter of this fit can be related to the position of the interaction sites within the target. The extracted slope parameter is 177$\pm$23 $(\text{GeV/c})^{-2}$, which is consistent with that expected for an Au nucleus (199 $(\text{GeV/c})^{-2}$)~\cite{UPC_PT,PhysRevC.60.014903,KLEIN2017258} within uncertainties. As shown in the figure the data point at $-t <$ 0.001 $(\text{GeV/c})^{2}$ is significantly lower than the extrapolation of the exponential fit. This suppression may be an indication of interference, which has been confirmed by STAR~\cite{PhysRevLett.102.112301} in the UPC case. The theoretical calculation with interference from~\cite{PhysRevC.97.044910}, shown as the blue curve in the plot, can describe the Au+Au data reasonably well ($\chi^{2}/NDF$ = 4.8/4) for $-t < 0.015$ $($GeV$/c)^{-2}$. It should be aware that there also exists possible contribution from incoherent $J/\psi$ photoproduction. The fitting $-t$ range is chosen to ensure that the coherent production is dominant over
the incoherent production. Due to the different nuclear profile, the $-t$ distribution in U+U collisions is expected to be different from that in Au+Au collisions, however, as shown in the figure, the difference is not observed due to the large uncertainties. We would like to point out that the probability of a random coincidence of a minimum bias event with the coherent production of a J/$\psi$ in a UPC event in the same bunch crossing was found to be negligible. In the overall data sample, only 0.2 J$/\psi$ events from the random coincidence are expected for the full centrality range with the STAR detector acceptance and efficiencies.

\renewcommand{\floatpagefraction}{0.75}
\begin{figure}[htbp]
\begin{center}
\includegraphics[keepaspectratio,width=0.45\textwidth]{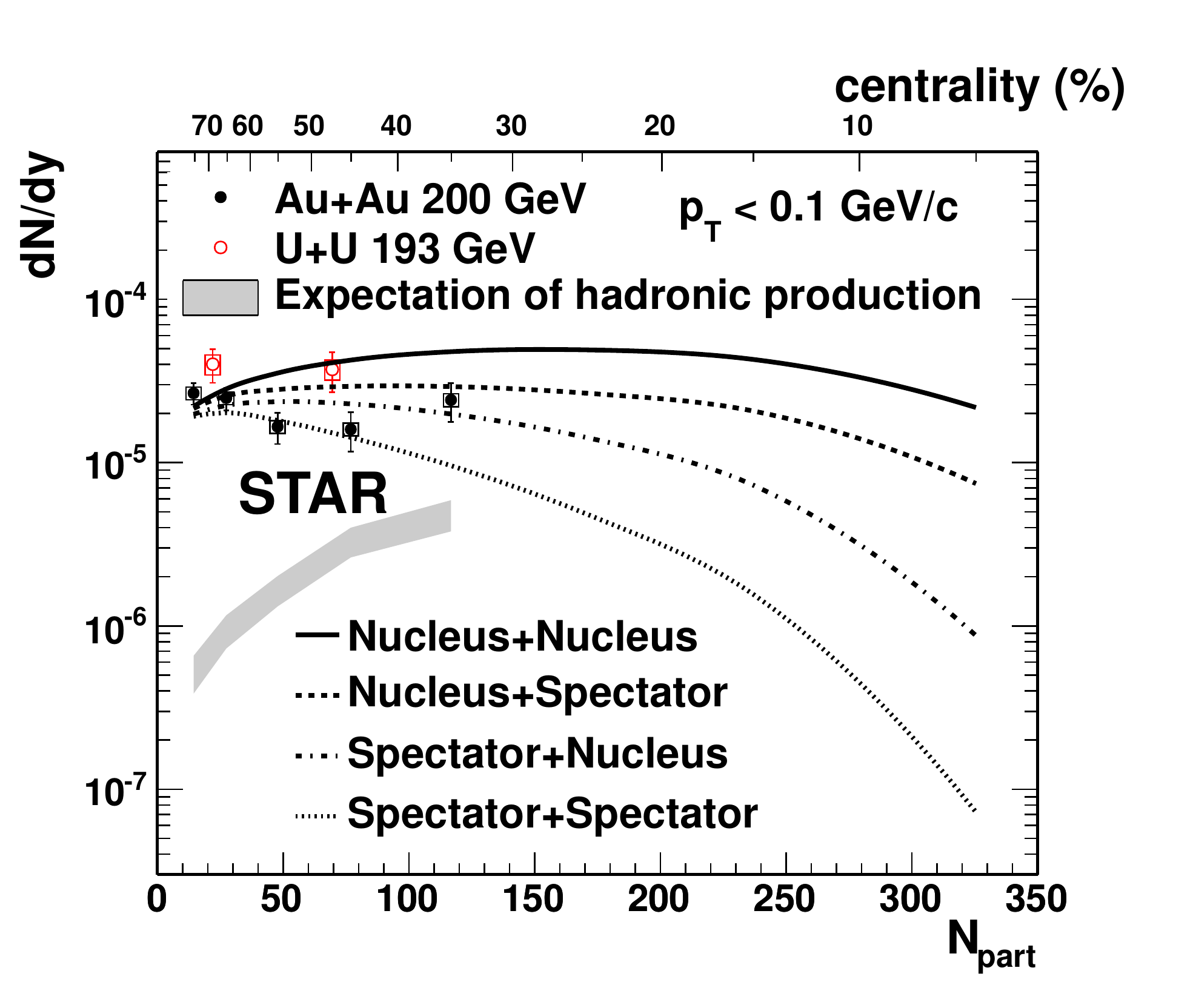}
\caption{(color online) The $p_{T}$-integrated J/$\psi$ yields ($p_{T} <$ 0.1~GeV/c) with the expected hadronic contribution subtracted as a function of $N_{\rm{part}}$ for 30-80$\%$ Au+Au collisions and 40-80$\%$ U+U collisions. The expected hadronic contributions for Au+Au collisions are also shown. The error bars represent the statistical uncertainties, while the boxes represent the systematic uncertainties. The lines are model calculations of coherent photoproduction with different scenarios for photon and Pomeron emitters~\cite{PhysRevC.97.044910}.}
\label{figure5}
\end{center}
\end{figure}

 Figure~\ref{figure5} shows  $p_{T}$-integrated J/$\psi$ yields for $p_{T} <$ 0.1 GeV/c with the expected hadronic contribution subtracted as a function of $N_{\rm{part}}$ for 30-80$\%$ Au+Au and 40-80$\%$ U+U collisions. The expected hadronic contributions in Au+Au collisions, extracted from the fit extrapolations in Fig.~\ref{figure2}, are also plotted for comparison. As depicted in the figure, the contribution from hadronic production is not dominant for the low-$p_{T}$ range in the measured centrality classes. Furthermore, the hadronic contribution increases dramatically toward central collisions, while the measured excess shows no sign of significant centrality dependence within uncertainties. Assuming that coherent photoproduction causes the excess at the very low $p_{T}$, the excess in U+U collisions should be larger than that in Au+Au collisions. Indeed the central value of measurements in U+U collisions is larger than that in Au+Au collisions. However, limited by the current measurement precision, the observed difference is not significant. The model calculations for Au+Au collisions with the coherent photoproduction assumption~\cite{PhysRevC.97.044910} are also plotted for comparison. In the model calculations, the authors consider either the whole nucleus or only the spectator nucleons as photon and Pomeron emitters, resulting in four configuration for photon emitter + Pomeron emitter:(1) Nucleus + Nucleus; (2) Nucleus + Spectator; (3) Spectator + Nucleus; (4) Spectator + Spectator. All four scenarios can describe the data points in the most peripheral centrality bins (60-80$\%$). However, in more central collisions, the Nucleus + Nucleus scenario significantly overestimates the data, which suggests that there may exist a partial disruption of the coherent production by the violent hadronic interactions in the overlapping region. The measurements in semi-central collisions seem to favor the Nucleus + Spectator or Spectator + Nucleus scenarios. The approach used in the model effectively incorporates the shadowing effect, which can describe the UPC results in the $x$-range probed by the RHIC measurement. However, the coherently produced J$/\psi$ could be modified by hot medium effects, e.g. color screening, which is not included in the model. More precise measurements toward central collisions and advanced modeling with hot medium effects included are
essential to distinguish the different scenarios.

In summary, we report on the recent STAR measurements of J/$\psi$ production at very low $p_{T}$ in hadronic Au+Au collisions at $\sqrt{s_{\rm{NN}}} =$ 200 GeV and U+U collisions at $\sqrt{s_{\rm{NN}}}=$ 193 GeV at mid-rapidity. Dramatic enhancements of yields are observed for $p_{T} <$ 0.2 GeV/c in peripheral collisions (40-80$\%$) beyond the conventional hadronic production modified by cold and hot medium effects. The observed excess shows no centrality dependence within uncertainties. In particular, the $dN/dt$ distribution in the very low $p_{T}$ range is presented for the first time, and shows apparent similarity to that of coherently produced vector mesons in ultra-peripheral collisions. The slope parameter extracted from the distribution is consistent with that expected for a Au nucleus, and an indication of interference is seen at the lowest $t$ values. Furthermore, theoretical calculations of coherent photoproduction can describe the excess yield in the most peripheral centrality class (60-80$\%$) reasonably well. On the other hand, the comparison between data and model calculations in semi-central collisions reveals that the coherent production may be partially disrupted by the concurrent hadronic interactions in the overlapping region. Based on the aforementioned observations, this strongly suggests that the significant excess observed at extremely low $p_{T}$ is likely to originate from coherent photoproduction in hadronic collisions. The coherently produced J/$\psi$'s in hadronic collisions may serve as an additional probe of QGP, and provide an opportunity to explore the gluon distribution in a nucleus. More differential measurements with better precision toward central collisions are called for in the future to better understand the origin of the low $p_T$ J$/\psi$ excess as well as to quantify its properties.

\textbf{Acknowledgement:} We thank the RHIC Operations Group and RCF at BNL, the NERSC Center at LBNL, and the Open Science Grid consortium for providing resources and support.  This work was supported in part by the Office of Nuclear Physics within the U.S. DOE Office of Science, the U.S. National Science Foundation, the Ministry of Education and Science of the Russian Federation, National Natural Science Foundation of China, Chinese Academy of Science, the Ministry of Science and Technology of China and the Chinese Ministry of Education, the National Research Foundation of Korea, Czech Science Foundation and Ministry of Education, Youth and Sports of the Czech Republic, Department of Atomic Energy and Department of Science and Technology of the Government of India, the National Science Centre of Poland, the Ministry  of Science, Education and Sports of the Republic of Croatia, RosAtom of Russia and German Bundesministerium fur Bildung, Wissenschaft, Forschung and Technologie (BMBF) and the Helmholtz Association.
\nocite{*}
\bibliographystyle{aipnum4-1}
\bibliography{aps}
\end{document}